\documentclass[nonacm]{acmart}

\usepackage{algpseudocode}
\usepackage{csquotes}
\usepackage{glossaries}
\glsdisablehyper %
\usepackage[inline-per-mode=symbol]{siunitx}
\newcommand{\cref}[1]{\autoref{#1}}
\newcommand{\Cref}[1]{\autoref{#1}}

\usepackage{subcaption}

\usepackage{todonotes}
\setuptodonotes{inline, noinlinepar}

\usepackage{tikz}
\usetikzlibrary{decorations.pathreplacing}

\usepackage{listings}
\lstalias{cpp}{c++}

\newacronym{os}{OS}{operating system}
\newacronym{cas}{CAS}{compare and swap}
\newacronym{sqs}{CES}{Combine-and-Exchange Scheduling}

\newcommand{\ie}{i.e.,\ }
\newcommand{\eg}{e.g.,\ }
\newcommand{\cf}{cf.\ }

\begin{document}

\title{Minimize Your Critical Path with Com\-bine-and-Exchange Locks}

\author{Simon K\"{o}nig}
\email{simon.koenig@ipvs.uni-stuttgart.de}
\affiliation{%
  \institution{University of Stuttgart}
  \country{Germany}
}
\author{Lukas Epple}
\email{lukas.epple@ipvs.uni-stuttgart.de}
\affiliation{%
  \institution{University of Stuttgart}
  \country{Germany}
}
\author{Christian Becker}
\email{christian.becker@ipvs.uni-stuttgart.de}
\affiliation{%
  \institution{University of Stuttgart}
  \country{Germany}
}

\renewcommand\footnotetextcopyrightpermission[1]{}
\settopmatter{printfolios=false}
\setcopyright{none}
\acmConference{}{}{}

\begin{abstract}
  
Coroutines are experiencing a renaissance as many modern programming languages support the use of cooperative multitasking for highly parallel or asynchronous applications.
One of the greatest advantages of this is that concurrency and synchronization is manged entirely in the userspace, omitting heavy-weight system calls.
However, we find that state-of-the-art userspace synchronization primitives approach synchronization in the userspace from the perspective of kernel-level scheduling.
This introduces unnecessary delays on the critical path of the application, limiting throughput.

In this paper, we re-think synchronization for tasks that are scheduled entirely in the userspace (\eg coroutines, fibers, etc.).
We develop \gls{sqs}, a novel scheduling approach that ensures contended critical sections stay on the same thread of execution while parallelizable work is evenly spread across the remaining threads.
We show that our approach can be applied to many existing languages and libraries, resulting in 3-fold performance improvements in application benchmarks as well as 8-fold performance improvements in microbenchmarks.

\glsresetall

\end{abstract}

\glsresetall

\maketitle

\glsresetall
\section{Introduction}

In general, userspace scheduling (\ie cooperative scheduling) has lower overhead compared to (preemptive) \gls{os} thread scheduling~\cite{Pufek2020,Karsten2020,Beronic2021,Seo1998,Seo1999}.
This is because userspace scheduling does not require a system call to exchange the currently running task (\ie fiber, coroutine, etc.).
Because of this, the industry is recently leading towards the usage of light-weight cooperatively scheduled tasks such as coroutines~\cite{kotlinx_coroutines,java_virtual_threads,smith_structured_concurrency,Adya2002}.
Some modern languages are built around userspace scheduling from the start~\cite{chakraborty_goroutines_2020}, many others have introduced support later in their evolution~\cite{nishanov_p0913r0_2018,java_virtual_threads,Elizarov2021,rust_tokio}.
Around these languages, there exist several frameworks that facilitate the execution of cooperatively scheduled tasks along with their scheduling and synchronization (\eg boost fibers~\cite{boost_fibers}, Java's \emph{Virtual Threads}~\cite{java_loom_proposal,java_virtual_threads}, Lua's coroutines~\cite{Ierusalimschy2007}, Kotlin coroutines~\cite{Elizarov2021,kotlinx_coroutines}, or Rust Tokio~\cite{rust_tokio}).
Furthermore, since the standardization of C++ coroutines~\cite{nishanov_n4780_2017}, many execution frameworks emerged~\cite{saelensminde_felsparcoro_nodate,baldwin_jbaldwinlibcoro_nodate,haim_david-haimconcurrencpp_nodate,kelbon_kelbonkelcoro_nodate,netcan_netcanasyncio_nodate,boost_cobalt,google_coros_cppnow,cppcoro,facebook_folly}.
Noteworthy examples are cppcoro \cite{cppcoro} and Facebook's folly library \cite{facebook_folly}.

The specific implementations and concepts differ largely across languages and libraries.
However, we find that none of these implementations fully utilize the possibilities of cooperatively scheduled tasks.
While they all correctly ensure threads do not block or busy-wait, they still come with one of the following shortcomings:
First, those frameworks that eagerly execute critical sections prevent the parallel execution of program parts outside the critical section~\cite{cppcoro,chakraborty_goroutines_2020,facebook_folly,baldwin_jbaldwinlibcoro_nodate,google_coros_cppnow}.
This causes, in the worst case, parallel execution to collapse to a single thread after tasks contended for a shared mutex.
Our evaluation confirms this shortcoming.
Second, other frameworks re-use the paradigms of thread-level synchronization in the context of userspace tasks.
Specifically, they lazily execute critical sections by waking the next waiter when a task leaves the critical section (\ie calling \texttt{notify()})~\cite{boost_fibers,rust_tokio,java_loom_proposal,java_virtual_threads}.
Hence, a thread that leaves the critical section notifies the next waiter and continues the execution of a parallelizable program section.
This way, the waiter does not enter the critical section right away but has to wait in a thread's ready queue.
This lock design prefers the execution of parallelizable program parts over the execution of the critical section.
Our evaluation shows that his decision prolongs the critical path of the application.
Queuing times are often an order of magnitude longer than the critical section itself.

These shortcomings are due to the fact that state-of-the-art implementations do not separate the critical section from the code immediately following the critical section.
In this paper, we present a novel approach that ensures that critical sections, and only the critical sections, of different tasks execute eagerly on a single thread while parallelizable (\ie non-critical) sections are deferred to other threads.
Moving the critical section to a specific thread is known as delegation-style locking~\cite{rcl, tclocks, sridharan2006thread, Hendler2010, ffwd}.
Delegation-style locks avoid the movement of shared data, thus improving cache locality.
To the best of our knowledge, this is the first paper to apply a delegation-style locking mechanism to the cooperatively scheduled context.
Our design, called \Gls{sqs}, applies two intuitive imperatives to the design of synchronization primitives:
Keep parallelizable program parts parallel and execute sequential parts as soon as possible.
For this, our approach transparently delegates critical sections and solves the two problems mentioned above: we minimize the length of the critical path while maximizing the parallel system utilization.

In short, the contributions of this paper are as follows:
\begin{itemize}
    \item We present \gls{sqs}, a new scheduling mechanism for critical sections that builds on top of the findings of delegation-style locks and rethinks synchronization in the userspace.
    \Gls{sqs} minimizes the length of the critical path and maximizes parallelism.

    \item We evaluate \gls{sqs} on NUMA machines and show that it improves the performance within and across sockets.
    In our benchmarks, we improve throughput by \SI{3.3}{\times} in application benchmarks and \SI{8.1}{\times} in microbenchmarks.
\end{itemize}

The remainder of this paper is structured as follows:
\cref{sec:related_work} introduces related work.
\Cref{sec:background} introduces the underlying concepts of synchronization primitives that use cooperative scheduling.
In \cref{sec:problem_statement}, we outline the problems incurred by state-of-the-art implementations.
In \cref{sec:our_scheduler}, we introduce the design and implementation of our approach.
\cref{sec:evaluation} presents our evaluations and measurements.
\cref{sec:conclusion} concludes the paper.

\section{Related Work}
\label{sec:related_work}

Parallel execution requires synchronization of shared data which can have adverse effects and reduce an application's throughput.
This scalability problem is at the center of over thirty years of research on the design of efficient and scalable lock implementations~\cite{chabbi2016contention, chabbi2015high, tclocks, luchangco2006hierarchical, magnusson1994queue, scott2001scalable, agarwal1989adaptive,  bacon2004thin, boguslavsky1994optimal, he2005preemption, mellor1991algorithms}.
The most widely used lock implementations move the ownership of the critical section between threads~\cite{mellor1991algorithms, dice2019compact, kashyap2019scalable}.
More recent research explores the opposite philosophy and moves the execution of the critical section to the thread that last owned the lock.
In this so-called delegation-style locking, a single thread of execution (often called the combiner) executes all critical sections~\cite{ffwd, sridharan2006thread, rcl, Hendler2010, tclocks}.

Generally, there are two categories of delegation-style synchronization:
Earlier approaches modified data structures to use delegation internally~\cite{Hendler2010, Oyama1999, ffwd}.
More recently, delegation is integrated into the implementation of lock synchronization primitives~\cite{rcl, tclocks}.
Delegation-style data structures modify the implementation of the data structure such that concurrent access is internally synchronized through the use of a combiner.
This approach is most commonly known as Flat Combining~\cite{Hendler2010}.
Because Flat Combining requires modifications of the data structure interface, it can be difficult to integrate into existing applications.

To avoid modifications of the data structure, other approaches integrate delegation into synchronization primitives such as locks.
Earlier implementations (\eg RCL~\cite{rcl}) still require modifications of the application code because delegation required the critical section to be captured by a function pointer.
TCLocks~\cite{tclocks} do not require such modifications because they implicitly transfer the critical section.
The awaiting thread creates an execution context out of its own stack and register state and transfers this to the combiner thread.
This way, TCLocks performs an implicit \gls{os} thread-level context switch in the userspace.

Furthermore, the approaches differ in the selection of the combiner thread.
In TCLocks~\cite{tclocks} or Flat Combining~\cite{Hendler2010}, the first thread to enter a contended critical section becomes the combiner.
In this case, the combiner thread is busy executing other critical sections, preventing it from making progress outside of the critical section.
Thus, the combiner thread experiences a higher critical section latency as it executes many other critical sections before returning.
The critical section latency is the time from calling \verb!lock()! until \verb!unlock()! returns.
Dynamic selection of the combiner also reduces performance in cases of low contention~\cite{rcl, Hendler2010}.
In contrast to that, RCL~\cite{rcl} and ffwd~\cite{ffwd} use a dedicated static combiner thread per data structure to achieve near constant critical section latency.
However, using one thread per data structure increases the overall system load and can lead to oversubscription (\ie more threads than available cores).
\Gls{sqs} implicitly selects one of the contending threads as the thread that executes all critical sections until contention drops.
Thus, only contended locks use delegation, in cases of low contention our implementation is equivalent to a spinlock.

Additionally, the waiting strategy while the combiner executes the critical section on behalf of the waiter impacts the execution of other work.
The goal of any strategy is to maximize the time spent doing useful work.
Existing implementations use a combination of busy-waiting and thread-level blocking~\cite{tclocks, Hendler2010, rcl, ffwd}.
This wastes CPU cycles in spin-loops or system calls.

Existing delegation-style locks experience difficulties in the actual implementation of delegation.
A major challenge is the transfer of the critical section to another thread of execution.
Implementations try to avoid system calls, as their overhead would outweigh the benefits of delegation.
Instead, implementations use some form of userspace scheduling to transfer control between threads of execution.
Specifically, the state-of-the-art either explicitly encapsulates the critical section as a function pointer~\cite{ffwd, rcl} or it relies on CPU architecture-specific instructions that implicitly transfer call stacks between \gls{os} threads~\cite{tclocks}.
Explicit encapsulation (\eg as done in ffwd \cite{ffwd} or RCL \cite{rcl}) has the downside that it requires rewriting large parts of the application and reduces the readability of code.
Existing implicit approaches such as TCLocks~\cite{tclocks} circumvent parts of the \gls{os} scheduler to implicitly transfer call stacks between threads.
This requires careful consideration of the kernel scheduling behavior and potentially kernel modifications.
In this paper, we highlight the following insight: Cooperatively scheduled tasks can transparently encapsulate the critical section without additional assumptions on the behavior of the \gls{os} scheduler.
Hence, a large portion of the complexities can be avoided through the usage of cooperatively scheduled tasks.

\section{Background and System Model}
\label{sec:background}

In this section, we introduce our terminology and explain state-of-the-art designs for synchronization primitives that rely on cooperative scheduling.

\subsection{Terminology}
This paper discusses cooperatively scheduled environments.
We call a \emph{task} the basic unit of concurrency.
A \emph{task} is a computation that has to be performed.
Furthermore, a task is cooperatively scheduled and, therefore, has the ability to explicitly yield control.
Each time the control flow reenters the task, the task resumes execution where it left off with its local control and data state retained (this is analogous to coroutines as described \eg by \cite{DeMoura2009}).
A \emph{thread} is the kernel-level concurrency abstraction.
A thread can be blocked and preempted by the \gls{os}.
A task has to be executed on a thread.
The execution of a task can be migrated to another thread whenever the task is suspended.
We call a set of threads that manage and execute tasks the \emph{executor}.
Parallelism can be achieved by running more than one thread in parallel at the \gls{os}-level.
We assume there to be a global executor known to the synchronization primitives.

Each application can be separated into parallelizable and sequential (\ie non-parallelizable) parts.
The speedup of a parallel application is bound by how much of the computation has to be executed sequentially~\cite{Amdahl1967}.
Therefore, a reasonable lower bound for an application's execution time is the consecutive execution of all sequential parts with no delays (the critical path).
In many applications, the parts of the computation that are difficult or even impossible to parallelize are those involving inter-thread communication or access to shared data structures~\cite{Hendler2010}.
For example, critical sections are such non-parallelizable parts.

\subsection{Task-Aware Mutex}

In cooperatively scheduled environments, one thread can execute thousands of tasks concurrently \cite{chakraborty_goroutines_2020,kohlhoff_asio_nodate,rust_tokio}.
Therefore, \gls{os}-level blocking is undesirable as this also blocks all other tasks on the same thread, analogous to a thread blocking anomaly \cite{Seo1999,Seo1998}.
Hence, applications avoid thread-level synchronization primitives in their tasks, and suspend tasks instead of blocking threads \cite{cppcoro,boost_fibers,kotlinx_coroutines}.
This is achieved through task-aware synchronization primitives.

In this section, we introduce a task-aware mutex and discuss the behavior of its \lstinline!lock! and \lstinline!unlock! operations.
We discuss other synchronization primitives in \cref{sec:other_sync_primitives}.
From here on out, we refer to a task-aware mutex as only \emph{mutex}.
Whenever we refer to a traditional blocking (\ie thread-level) mutex, we will state that explicitly.

The \lstinline!lock()! operation (\cf \autoref{fig:task_aware_mutex_lock}) first checks the state of the mutex.
If the mutex is in the UNLOCKED state, the locking task may continue into the critical section, we say it becomes the \emph{lock owner}.
Otherwise the task becomes a \emph{waiter} for the mutex.
In that case, the task suspends and its continuation is appended to the queue of waiters.
When the current lock owner leaves the critical section, it ensures that one of the waiters resumes, thus, passing the ownership of the lock to the next waiter.
In contrast to kernel-level scheduling, a cooperative mutex implementation selects the next lock owner explicitly.
When a task calls \lstinline!unlock()!, it passes the lock ownership to exactly one of the waiters.
From this point forward, only this task can enter the critical section until it calls \lstinline!unlock()! again.
Importantly, while the lock experiences contention, it remains in the locked state the entire time and no other task can \enquote{overtake} the next selected waiter.

\begin{figure}
    \begin{lstlisting}[language={c++}, gobble=6]
      void mutex::lock(current_continuation) {
        if(mutex.state == UNLOCKED) {
          mutex.state = LOCKED;
          current_continuation.resume();
        } else {
          mutex.waiters.push_back(current_continuation);
          // we have become a waiter
        }
      }
    \end{lstlisting}
    \caption{
      Pseudo code of the lock method of a task-aware mutex.
      For simplicity, this pseudo code assumes the entire method call is atomic.
      Implementations use lock-free or locking methods to achieve this.
      }
    \Description{}
    \label{fig:task_aware_mutex_lock}
\end{figure}

There exist two different strategies for this transfer of ownership in the state-of-the-art:
\emph{Inline schedulers} resume a waiting task on the currently active thread while blocking the previous task.
\emph{Dispatch schedulers} notify a waiting task that it can be resumed on another thread.
The following paragraphs present the behavior of these two approaches in more detail.

\paragraph{Inline Schedulers}

The first design resumes the next waiter directly inside the call to \lstinline{unlock} (\cf \autoref{fig:inline_scheduling_decision}).
Note that the unlocking task does not suspend, because resuming the next waiter behaves like a regular function call.
This behavior repeats recursively until the queue of waiters is empty.
Hence, an unlocking task returns from its call to \lstinline{unlock} only after all following waiters have passed the critical section as well.
We refer to this type of design as an \emph{inline scheduler}.
State-of-the-art implementations that follow this design include cppcoro~\cite{cppcoro}, Go~\cite{chakraborty_goroutines_2020}, facebook's folly~\cite{facebook_folly} and Google's coroutine library K3~\cite{google_coros_cppnow}.

\begin{figure}
    \begin{lstlisting}[language={c++}, gobble=6]
      void inline_mutex::unlock() {
        if(mutex.waiters.empty()) {
          mutex.state = UNLOCKED;
        } else {
          waiter = mutex.waiters.pop_front();
          // Execute next waiter inline
          waiter.resume();
        }
      }
    \end{lstlisting}
    \caption{
      Pseudo code of the unlock method of an inline scheduling mutex.
      Note that the unlocking task does not suspend and, therefore, cannot resume on another thread!
      The current task's call to unlock returns only after the waiter suspends from the nested resume call.
      }
    \Description{}
    \label{fig:inline_scheduling_decision}
\end{figure}

\paragraph{Dispatch Schedulers}

The second design approaches the problem from the point of view of kernel-level synchronization.
In particular, it transfers ownership of the mutex by moving the next waiting task's continuation to the ready queue of the executor (\cf \autoref{fig:runtime_scheduling_decision}).
The resulting behavior is equivalent to \lstinline{notify()}-ing a blocked thread at the OS-level.
We refer to this design as \emph{dispatch scheduling}.
State-of-the-art implementations that follow this design include Kotlin's coroutines~\cite{kotlinx_coroutines}, Boost fibers~\cite{boost_fibers} and Rust's Tokio~\cite{rust_tokio}.

\begin{figure}
    \begin{lstlisting}[language={c++}, gobble=6]
      void dispatch_mutex::unlock() {
        if(mutex.waiters.empty()) {
          mutex.state = UNLOCKED;
        } else {
          waiter = mutex.waiters.pop_front();
          // Dispatch next waiter
          executor.schedule(waiter);
        }
      }
    \end{lstlisting}
    \caption{
      Pseudo code of the unlock method of a dispatch scheduling mutex.
      The current task's call to unlock returns immediately.
      However, the next waiter waits in the ready queue of the executor while no other task can enter the critical section (the mutex state is LOCKED).
      }
    \Description{}
    \label{fig:runtime_scheduling_decision}
\end{figure}

\subsection{Delegation-Style Locks}

The idea of delegating the critical section to another thread instead of transferring the ownership of the protected resource is called delegation-style locking.
With a delegation-style lock, the critical sections of many waiters eventually execute on the same thread, called the \emph{combiner}.
State-of-the-art implementations of this type of lock design include TCLocks and RCL~\cite{tclocks, rcl}.
Both these implementations require \gls{os}-level threads as the unit of concurrency.
To implement the delegation of the critical section, these kernel-level implementations require a form of task scheduling in the userspace.
For example, a waiter in a TCLock creates an execution context out of its own stack and register state and transfers this to the combiner thread.
This way, TCLocks implicitly performs a (cooperative) thread-level context switch.

Inline scheduling is a form of delegation-style locking because the unlocking thread resumes the continuation of the next lock waiter.
To the best of our knowledge, inline scheduling is the only delegation-style lock implementation that works in a fully cooperatively scheduled environment.

\section{Problem Statement}
\label{sec:problem_statement}

It is a wide-spread intuition that a high processor utilization must also corresponds to high throughput, as long as threads do not busy-wait.
The reasoning behind this is that all threads are fully utilized, thus, are getting work done.
In particular, it is okay for tasks to wait in the ready queue as long as the thread is still getting work done.

With this paper, we aim to highlight a problem that hides in the details of this intuition.
We aim to show that in circumstances that require queuing, it is beneficial (in terms of throughput) to prioritize specific types of work, and prevent them from ever having to wait in a ready queue.
Tasks executing sequential parts of the program (i.e., critical sections) should not be delayed and their execution should have priority over parallelizable work.
Our justification for this has its basis in the concept of the application's critical path.
Forcing tasks that execute critical sections to wait in a ready queue delays the execution of a segment of the critical path.
Hence, we prolong the critical path overall, reducing throughput~\cite{Amdahl1967}.

We clarify this problem in the following section.

\subsection{Queuing Delay}

\begin{figure}
    \centering
    \includegraphics{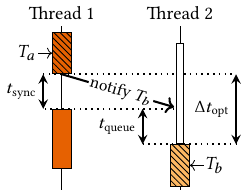}
    \caption{Delays incurred by dispatch schedulers at critical section boundaries. Patterned regions are critical sections, non-patterned regions are parallelizable.}
    \Description{A task (T a) leaves the critical section. It then notifies (T b) such that it can resume on another thread of execution. The time it takes to notify is the synchronization delay. The time after (T b) being notified and it starting its execution is the queuing delay. The sum of synchronization delay and queuing delay is delta t opt.}
    \label{fig:delays}
\end{figure}

We define \emph{queuing delay} ($t_{\mathrm{queue}}$) as the time during which no task is in the critical section even though at least one task is waiting to enter.
In other words, it is the time between a waiter being scheduled for execution and the actual start of its execution in the critical section.
\Cref{fig:delays} shows the queuing delay for the task $T_b$, which is the next waiter after $T_a$.
$T_b$ is scheduled to run on thread 2.
As thread 2 is currently busy, $T_b$ has to wait in the ready queue of thread 2.
Both the synchronization delay $t_{\mathrm{sync}}$ and the queuing delay $t_{\mathrm{queue}}$ comprise overhead.
We denote the sum of the two delays as $\Delta t_{\mathrm{opt}}$.
Note, that there is no task in the critical section (visualized with pattern) in the duration $\Delta t_{\mathrm{opt}}$.
Thus, reducing $\Delta t_{\mathrm{opt}}$ has a direct influence on the program speedup under multi-threaded execution.

However, previous work mostly disregards the queuing delay as it does not clearly differentiate between $t_{\mathrm{sync}}$ and $\Delta t_{\mathrm{opt}}$.
To the best of our knowledge, queuing delay in cooperatively scheduled environments is only addressed through the use of work stealing in thread pools (\eg \cite{Blumofe1999,Lin2020,Tzannes2010}).
However, work stealing does not eliminate the queuing delay, it may only reduce it.
Queuing delay grasps a behavioral symptom of state-of-the-art task-based synchronization designs.
We will now discuss how both existing designs, \ie dispatch scheduling and inline scheduling, either directly or indirectly incur queuing delay on the critical path.
Hence, the synchronization primitive's design limits the overall throughput of the application.

\begin{figure}[t]
    \begin{subfigure}[t]{\linewidth}
        \centering
        \includegraphics{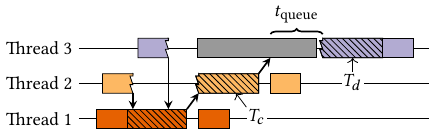}
        \caption{Dispatch scheduling}
        \Description{An unlocking task runs on Thread 1. It schedules the resumption of task (T c) on Thread 2. (T c) then schedules the resumption of (T d) on Thread 3. Since Thread 3 is busy at this time, the scheduled tasks remains in the queue until Thread 3 is ready. This is the cause of queuing delay.}
        \label{fig:runtime_scheduling}
    \end{subfigure}

    \vspace*{1em}

    \begin{subfigure}[t]{\linewidth}
        \centering
        \includegraphics{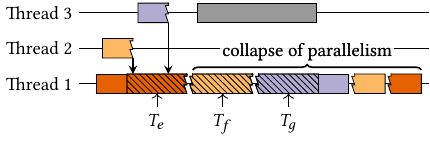}
        \caption{Inline scheduling}
        \Description{Task (T e) resumes (T f) inline. This recursively resumes task (T g). Since all tasks resume inline, they all run on Thread 1, collapsing parallelism.}
        \label{fig:inline_scheduling}
    \end{subfigure}
    \caption{Three tasks on three threads of execution.
    Each color represents one task, patterned regions indicate that the task is currently inside the critical section.
    }
    \label{fig:inline_and_runtime_scheduling}
\end{figure}

In dispatch scheduling, a waiter is not immediately resumed but scheduled for execution on the executor.
This does not guarantee that the waiting task runs immediately if all executor threads are busy at the moment.
\Cref{fig:runtime_scheduling} shows an example schedule produced by a dispatch scheduler.
The task $T_c$ leaves the critical section and notifies $T_d$.
However, $T_d$ has to wait in thread 3's ready queue while another task occupies the thread.
This incurs a queuing delay and creates a gap between the two sequential parts.

Traditionally, implementations apply work-stealing to shorten such waiting times in the ready queue.
However, work-stealing can only reduce the effects of queuing delay, but not circumvent it entirely.
A thief can only steal the task if it is inside the queue, \ie already incurring queuing delay.
The goal of this paper is to eliminate queuing delay for tasks that are about to enter a critical section.

\subsection{Collapse of Parallelism}

The second possible design is inline scheduling, which initially eliminates queuing delay by executing the next critical section inside the call to \lstinline!unlock()!.
However, because the resumption of the waiter prevents the unlocking task to continue, this implementation limits parallelism.
The important part is, that to re-gain parallelism in an inline scheduling system, we again have to incur queuing delay on the critical path.
In this section, we will first show how inline scheduling limits the effective use of parallelism and then show how the solution to this problem incurs queuing delay.

\Cref{fig:inline_scheduling} shows an example schedule produced by an inline scheduler.
When task $T_e$ releases the mutex, its thread resumes $T_f$.
Note, that this resumption behaves exactly like a regular function call.
The problem here is not queuing delay but the fact that $T_e$ has to wait despite the fact that it no longer executes a sequential part and could run parallel to $T_f$.
The same is true for $T_f$, which resumes $T_g$ after leaving the critical section.
Finally, $T_g$ does not find any waiters for the mutex and continues after its critical section.
When $T_g$ finishes its execution completely, it returns control back to $T_f$.
Eventually, $T_f$ returns control to $T_e$.
The order of task completion is the inverse of their initiation because all tasks remained active on thread 1's call stack.

Thus, inline scheduling results in sequential execution of potentially parallelizable tasks.
A single thread will end up executing all tasks that contend for the mutex.
Therefore, we say parallelism collapsed down to a single thread.
\emph{Collapse of parallelism} is a fundamental problem of inline scheduling that cannot be circumvented by work-stealing.
This is because inline schedulers do not enqueue work on the executor but manage outstanding work on the thread's call stack, hence it is unavailable for thieves.
Implementations that incur collapse of parallelism include, \eg cppcoro~\cite{cppcoro} and facebook's folly library~\cite{facebook_folly}.

\paragraph{Indirect Queuing Delay}

\newcommand{\tunlock}{$T_e$}
\newcommand{\twaiting}{$T_f$}

The reasoning behind inline scheduling is that these synchronization primitive designs leave scheduling decisions up to the application~\cite{cppcoro,google_coros_cppnow}.
Hence, the synchronization primitive is not aware of an executor, \ie it cannot delegate the execution of a waiter to another thread.
Thus, the unlocking task \emph{has to} resume the waiting task inline, otherwise liveness would be violated.

Inline scheduling libraries introduce a scheduler concept to be used by the application (\eg \cite{cppcoro}).
The application code can call \lstinline!executor.schedule()! from inside a task.
This suspends the current task and enqueues it for resumption on the \lstinline!executor!.
To ensure that \twaiting{} and \tunlock{} run in parallel, the application has to call \lstinline!executor.schedule()! right after \twaiting{} gains control.
This suspends \twaiting{} and allows \tunlock{} to continue.
Eventually, the executor resumes \twaiting{} and the two tasks can run in parallel.
However, note that the application would need to place the call to \lstinline!executor.schedule()! \emph{inside} the critical section.
Hence, this solution exhibits the same behavior as a dispatch scheduler and incurs queuing delay on the critical path.

\section{Combine-and-Exchange Scheduling (CES)}
\label{sec:our_scheduler}

In this section, we present \gls{sqs}, a novel scheduling approach for synchronization primitives in cooperatively scheduled environments.
\Gls{sqs} aims to minimize the queuing delay for tasks that are about to enter a critical section.
To achieve this, it places the execution of successive critical sections on the same thread of execution.
Meanwhile, tasks that leave the critical section are moved to other threads, ensuring the system utilizes the available parallelism.

A mutex that implements \gls{sqs} uses the same \lstinline!lock()! method as any of the state-of-the-art mutex implementations (\cf \cref{fig:task_aware_mutex_lock}).
That is, when a task $T$ tries to \lstinline!lock! the mutex but another task is already in the critical section, we suspend $T$ and append it to the list of waiters.
Otherwise, $T$ enters the critical section immediately.
\Gls{sqs} differs from the state-of-the-art when a task $T$ calls \lstinline!unlock()! (\cf \cref{fig:sqs}).
First, we check if there are any waiters.
If yes, suspend $T$ and schedule $T$ on \emph{another} thread on the executor.
Immediately resume the first waiter on the current thread.
If there are no waiters, do not suspend $T$ and continue its execution.

\begin{figure}
  \begin{lstlisting}[language={c++}, gobble=6]
      void CES_mutex::unlock(current_continuation) {
        if(mutex.waiters.empty()) {
          mutex.state = UNLOCKED;
          current_continuation.resume();
        } else {
          waiter = mutex.waiters.pop_front();

          // Dispatch current task
          executor.schedule(current_continuation);
          // Immediately resume waiter
          waiter.resume();
        }
      }
    \end{lstlisting}
    \caption{
      Pseudo code of the unlock method of a \gls{sqs} mutex.
      The current task suspends, its continuation is dispatched on the executor whereas the next waiter resumes immediately on the current thread.
      }
    \Description{}
    \label{fig:sqs}
\end{figure}

State-of-the-art dispatch schedulers notify the next waiter and do not suspend the current mutex owner.
This decision incurs queuing delay.
In contrast, \gls{sqs} suspends the previous lock owner and immediately resumes the next waiter, eliminating queuing delay.
Compared to inline scheduling, \gls{sqs} allows the unlocking task to resume on another thread.
This prevents collapse of parallelism.

An additional benefit of \gls{sqs}'s design is in its improved cache locality at runtime.
\Gls{sqs} ensures that a task that is blocked because another task is holding the mutex will later execute on the same thread as the current mutex owner.
This way, we automatically execute all critical sections of a contended mutex in a single thread.
Hence, under high contention, the protected resource data may already be in a valid cache line, ready to be used by the new lock holder.
This also implies that modifications to the resource rarely have to cross thread boundaries.
Hence, \gls{sqs} reduces synchronization overheads and cache invalidations.

From an implementation point-of-view, \gls{sqs} achieves this by placing a second suspension point at the \lstinline!unlock! operation.
State-of-the-art implementations only conditionally suspend a task at the \lstinline!lock! operation.
With \gls{sqs}, \lstinline!unlock! operations may suspend the calling task too.
While this does introduce additional suspension points, it makes up for this increase in book-keeping overhead by collapsing contending resource access to a single thread, eliminating queuing delay and improving cache locality.

\paragraph{Assumptions}

During the design, our projected performance improvements of \gls{sqs} are based on two assumptions:
\begin{enumerate}
  \item The migration delay to move the code (\ie the execution of the critical section) is smaller than the delay incurred by moving the shared resource.
  \item The overhead of switching the current task (which includes a suspension) is smaller than the queuing delay.
\end{enumerate}
The first assumption is at the basis of all delegation-style locks, and has been shown to be true numerous times across many different applications~\cite{ffwd, rcl, Hendler2010, tclocks}.
The second assumption is new to this work.
Our evaluation shows that this is true in cooperatively scheduled environments where no thread idles.

\paragraph{Comparison to Delegation-Style Locks}

\Gls{sqs} is a dele\-gation-style lock design.
Waiters delegate the execution of their critical section to a single thread, often called the \emph{combiner}.
This improves performance due to increased cache locality.

\begin{figure}
    \begin{subfigure}[t]{\linewidth}
        \centering
        \includegraphics{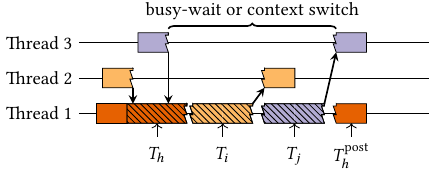}
        \caption{Delegation-Style Locks}
        \Description{Three tasks contend for a single mutex. Thread 1 executes all critical sections. Threads 2 and 3 delegate their critical sections to thread 1. After the critical section finished, the tasks return to their original thread.}
        \label{fig:delegation_locking}
    \end{subfigure}

    \vspace*{1em}

    \begin{subfigure}[t]{\linewidth}
        \centering
        \includegraphics{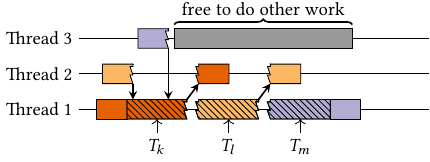}
        \caption{\gls{sqs}}
        \Description{Three tasks contend for a single mutex. Thread 1 executes all critical sections. Task Tk starts execution on thread 1. After the critical section finishes, this task migrates to a different free thread, thread 2 in this example. Threads 2 and 3 migrate their critical sections to the combiner, instead of blocking or buys-waiting, they are immediately available to execute other tasks.}
        \label{fig:our_scheduler}
    \end{subfigure}

    \caption{
        Scheduling behavior of delegation-style locks compared to our approach \gls{sqs}.
        Patterned regions represent the critical section of the tasks.
    }
    \Description{}
    \label{fig:delegation_vs_our_scheduler}
\end{figure}

Delegation-style locks have been extensively researched in preemptively scheduled environments (\eg~\cite{rcl, tclocks}).
However, to the best of our knowledge, this paper is the first to apply true delegation-style locking to the cooperatively scheduled context.
\Cref{fig:delegation_vs_our_scheduler} compares the behavior of state-of-the-art delegation-style locks with our approach.
Resource access occurs on a single thread in both cases.
Because \gls{sqs} works in a cooperatively scheduled context, we can more liberally move tasks between threads of execution.
State-of-the-art delegation-style locks require threads to busy-wait \cite{tclocks} or context switch \cite{rcl} while the combiner (Thread 1) executes the critical section on their behalf (\cf Thread 3 in \cref{fig:delegation_locking}).
In contrast, \gls{sqs} moves the execution of $T_m$ from thread 3 to thread 1 while thread 3 continues with other work (\cf \cref{fig:our_scheduler}).

Another issue with preemptive delegation-style locks is the critical section latency of the combiner.
Work that runs on the combiner's thread ($T_h$) is delayed while the combiner executes critical sections.
Note, that thread 1 executes $T_h^{\mathrm{post}}$ only after all contending critical sections have finished.
This delay can be arbitrarily long.
Moreover, the section $T_{h}^{\mathrm{post}}$ might also produce more contention on critical sections in the future.
With our \gls{sqs} mutex, $T_k$ which initially started on thread 1 (the combiner) was migrated to thread 2.
Thus, reducing the critical section latency.

\subsection{Implementation}

Both, the \lstinline!lock! and \lstinline!unlock! methods need to check the state of the mutex and interact with the queue of waiters atomically.
During implementation of \gls{sqs}, we experimented with the details of how to achieve this atomicity.
Our final implementation uses a test-and-set spin lock to protect the state variable and an intrinsic queue of waiters.
We have also experimented with a lock-free alternative where the state-variable takes three possible states, UNLOCKED, LOCKED\_NO\_WAITERS, and a pointer to the first waiter.
This design is inspired by cppcoro's handling of the waiter queue~\cite{cppcoro}.
However, in our evaluations, we found the final spin locked alternative to perform more reliably across our benchmarks.

\subsection{Synchronization Primitives Beyond Mutex}
\label{sec:other_sync_primitives}

Up to this point, we have only discussed the scheduling behavior of a task-aware mutex.
In this sub-section, we extend our previous discussion to other task-aware synchronization primitives.
We believe that this list covers the most important principles that, in a similar way, apply to any other synchronization primitive.

\paragraph{Semaphore}

The primary example is a general semaphore where one task may release more than one permit, possibly allowing multiple waiters to enter.
In a general semaphore, a single call to \mbox{\lstinline!acquire()!} or \mbox{\lstinline!release()!} can increment or decrement the available permits by more than one.
In this case, \mbox{\lstinline!release(n)!} suspends the calling task and schedules it in the same way as a mutex.
If $n>1$, it is possible that more than one waiter is now allowed to enter.
Hence, the semaphore resumes one waiter inline and schedules the resumption of up to $n-1$ waiters on the executor.
Thus, at least one waiter does not suffer queuing delay.

\paragraph{Reader-Writer Mutex}

We differentiate between the behavior of the two methods \lstinline!write_unlock! and \lstinline!read_unlock!.
The method \lstinline!read_unlock! checks whether the calling task is the only remaining reader.
If there are still other active readers, it continues the unlocking task.
However, if the calling task is the only remaining active reader, the method has to check whether there is a writer in the queue of waiters.
This writer is resumed inline, while the task calling \lstinline!read_unlock()! suspends and is scheduled on another thread.
A task calling \lstinline!write_unlock()! suspends and is scheduled to another thread if there are waiters in the queue.
If the next waiter is a writer, we resume this waiter inline.
Otherwise, we resume one reader inline and schedule the execution of all remaining readers.
Again, at least one waiter does not suffer queuing delay.

\paragraph{Condition Variable}

A condition variable supports the operations \lstinline!wait!, \lstinline!notify_one!, and \lstinline!notify_all!.
It always works in conjunction with a mutex.
When a task calls \lstinline!wait!, it unlocks the mutex, suspends and becomes a waiter until another task calls one of the \lstinline!notify! functions.
Calling \lstinline!notify! will wake one (or all) of the condition variable's waiters which will then try to reacquire the mutex.
If the mutex is locked while a task calls \lstinline!notify!, a waiter for the condition variable becomes a waiter for the mutex.
However, this process requires that the notified task resumes and then immediately suspends again.
Resuming a task for such a short duration has adverse effects on cache locality, thus, should be avoided when possible.
In practice, it is a common pattern to notify a condition variable while holding the associated mutex (\eg in LevelDB).
Hence, it is worth optimizing the interaction between condition variable and mutex.

In our implementation, calling \lstinline!notify! retrieves one (or all) of the waiters from the condition variable's waiter queue.
Then, without suspending the notifying task and without resuming the waiter(s), we test the availability of the mutex.
If the mutex is LOCKED, we append a new waiter.
Otherwise, we immediately resume the notified task, which now owns the mutex.

Optimizing the interaction between condition variable and mutex is already common in many standard implementations in preemptively scheduled environments.
However, in the preemptively scheduled context, calling notify still only schedules the notified thread for execution if the mutex is UNLOCKED.
With \gls{sqs}, we give the notified task priority over the notifying task because it owns a mutex and executes a critical section.
Hence, we transfer control to the notified task immediately, eliminating queuing delay.

\subsection{Implementation of a CES-based Coroutine Library}

A reference implementation of the concepts presented in this paper will be made publicly available upon acceptance of the paper.
We support standard C++ coroutines as the underlying asynchronous task type.
The library is built with \gls{sqs} as the core principle.
Based on this, we provide reference implementations for multiple synchronization primitives including (reader-writer) mutexes, channels, events, barriers, latches, etc.
As the underlying executor, the library uses a NUMA-aware work stealing approach with a queue implementation based on BWoS~\cite{Wang2023}.
We use this library in the following evaluations.
Furthermore, we provide a proof-of-concept implementation of \gls{sqs} for Java's Virtual Threads and boost fibers.
This shows that the scheduling concept of \gls{sqs} can be applied to many different frameworks, environments, and programming languages.

\subsection{Validation}

To confirm that \gls{sqs} keeps highly contended critical sections on the same thread, we tracked the executor thread IDs for a set of concurrent tasks.
Hereby, \num{1000} tasks are executed concurrently on \num{16} threads, each task repeatedly accesses one of four resources randomly.
This experiment showed that \gls{sqs} behaves as expected per our design, \ie, that the critical section of the \gls{sqs} mutex remains on a dedicated thread per resource as long as contention is high enough.
The execution of all tasks that try to access a locked mutex will be delegated to the thread of the current mutex owner.

The experiment also confirmed that \gls{sqs} only applies delegation when there is contention.
With reduced contention (increasing the number of resources), resource access is free to switch the executing thread.
This is in contrast to RCL~\cite{rcl} or ffwd~\cite{ffwd}, where resource access of a given resource may only happen on a specific core or thread.

\section{Evaluation}
\label{sec:evaluation}

\DeclareSIUnit\ops{ops}

In this section we evaluate \gls{sqs} by answering the following questions:
\begin{description}
    \item[Q1] How do the scheduling decisions of \gls{sqs} impact the performance of a mutex (\cref{sec:mutex_performance_comparison}) and reader-writer mutex (\cref{sec:rw_mutex_performance})?
    \item[Q2] Is \gls{sqs} generalizable to other languages and environments? (\cref{sec:sqs_in_other_langs})
    \item[Q3] Does \gls{sqs} improve the performance on the application-scale? (\cref{sec:application_benchmarks})
    \item[Q4] Are differences in performance due to queuing delay prolonging the critical path of the application? (\cref{sec:impact_queuing_delay})
\end{description}

\paragraph{Evaluation setup}

Our measurement setup comprises two machines.
First, we use a 2 socket AMD machine with 128 physical cores (max. 256 threads) that has \SI{2}{\tera\byte} of RAM split across the two NUMA nodes.
Second, we use an Intel machine with 64 physical cores split across 4 NUMA nodes with no hyperthreading (4 socket machine), it has \SI{1}{\tera\byte} of RAM.
Every benchmark has been compiled with gcc 14.2.1 and executed on Linux 6.8.0-53.
Benchmarks compiled with clang 18.1.3 produce similar results.
Measurement results are calculated as the the median of five benchmark runs.

\subsection{Mutex Microbenchmarks}
\label{sec:mutex_performance_comparison}

In this micro-benchmark, we evaluate performance differences between inline schedulers, dispatch schedulers and \gls{sqs} for a mutex.
\Cref{alg:mutex_benchmark_task} shows pseudo code for the task implementation that we used as part of our benchmark program.
A task consists of a loop, in which the task selects a random large prime, modifies a shared map inside the critical section and then calculates the sieve of Eratosthenes for the chosen number.
Hence, each task spends only a small fraction of time inside the critical section and a large fraction of time in parallelizable parts of the program.
Our benchmark program executes \num{5000} concurrent tasks.
Throughput is the number of completed loop cycles divided by the overall runtime.

\begin{figure}
  \centering
  \includegraphics{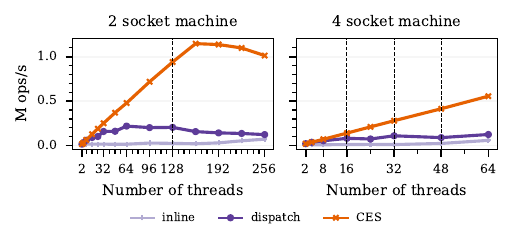}
  \caption{
        Impact of mutex scheduling decisions on throughput.
        Inline scheduling is equivalent to the implementation of the cppcoro~\cite{cppcoro} and folly~\cite{facebook_folly} libraries.
        }
  \Description{Throughput of inline scheduling remains constant across the number of threads on both machines. Dispatch scheduling suffers increased queuing delay, reducing throughput. Our approach continues to scale across NUMA nodes.}
  \label{fig:mutex}
\end{figure}

\Cref{fig:mutex} shows our measurement results.
As expected, inline scheduling incurs collapse of parallelism.
Because \gls{sqs} minimizes the movement of shared data, our scheduler continues to scale its throughput, considerably outperforming dispatch scheduling in a highly-parallel setting.
At 256 threads, \gls{sqs} achieves \SI{8.1}{\times} the throughput of dispatch scheduling on the two socket machine (\SI{0.124}{\mega\ops\per\second} and \SI{1.01}{\mega\ops\per\second} respectively).

\begin{figure}
  \begin{algorithmic}
      \Function{Task}{mutex, map}
        \For{1000 iterations}
          \State p $\overset\$\leftarrow$ PRIMES
          \State mutex.lock();
          \State map.insert(p, p)
          \State mutex.unlock();
          \State factors(p);
        \EndFor
      \EndFunction
  \end{algorithmic}
  \caption{Pseudo code of the benchmark task.}
  \Description{}
  \label{alg:mutex_benchmark_task}
\end{figure}

\paragraph{Influence of critical section length}

\begin{figure}
  \centering
  \includegraphics{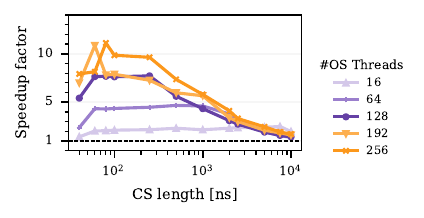}
  \caption{
  Speedup of \gls{sqs} over dispatch scheduling depending on the length of the critical section.
  A thread count larger than 128 spans both NUMA nodes.
  }
  \Description{Speedup of our approach is largest for small critical sections. With increasing critical section length, the speedup reduces. Increasing the number of threads used also increases the maximum speedup.}
  \label{fig:mutex_speedup}
\end{figure}

We now modify our benchmark program and increase the critical section length with varying amounts of busy-waiting.
\Cref{fig:mutex_speedup} shows the speedup of \gls{sqs} over dispatch scheduling.
As the previous measurement showed, \gls{sqs} generally scales better with the number of threads.
Up to \SI{250}{\nano\second} spent in the critical section, \gls{sqs} improves the throughput by one order of magnitude compared to dispatch scheduling on \num{256} threads.
For small critical sections, the queuing delay is large in relation to the time spent in the critical section.
Hence, in these cases, \gls{sqs} has the highest potential to improve overall throughput.

With increasing time spent in the critical section, the speedup decreases.
This is because the queuing delay contributes a smaller amount of overhead relative to the length of the critical section.
Note, that this is true for any form of delegation-style lock.
While the absolute improvements decrease, \gls{sqs} is still always faster than dispatch scheduling across all critical section lengths.
For a critical section that is \SI{7.5}{\micro\second} long, \gls{sqs} is still approximately \SI{2}{\times} faster than dispatch scheduling.
We have measured that \SI{95}{\%} of all critical sections in LevelDB are shorter than \SI{1}{\micro\second} (the median is below \SI{350}{\nano\second}), independent of the number of threads.
Hence, in real-world applications, \gls{sqs} has great potential to reduce the overhead of synchronization.

\subsection{Reader-Writer Mutex Microbenchmarks}
\label{sec:rw_mutex_performance}

\begin{figure}
    \centering
    \includegraphics{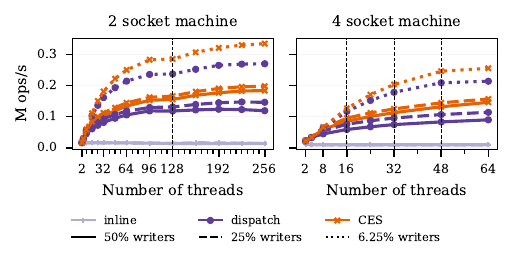}
    \caption{
        Impact of reader-writer mutex scheduling decisions on throughput.
        Inline scheduling is equivalent to the implementation of the cppcoro~\cite{cppcoro} and folly~\cite{facebook_folly} libraries.
    }
    \Description{Inline scheduling exhibits the smallest throughput independent of number of threads used. The writer percentage has no influence as well. Both our approach and dispatch scheduling scale with the number of threads including across NUMA nodes. However, our approach consistently outperforms dispatch scheduling across all measured configurations.}
    \label{fig:rwmutex}
\end{figure}

In this benchmark, we evaluate the performance of a fair reader-writer mutex implemented with either inline scheduling, dispatch scheduling or \gls{sqs}.
This compares the scheduling behavior in cases where more than waiter needs to be woken up (write access compared to read access).

\Cref{fig:rwmutex} shows the results of our measurements on both machines.
Inline scheduling experiences collapse of parallelism in all cases.
Both \gls{sqs} and dispatch scheduling scale with the number of threads.
Decreasing the writer percentage improves throughput because larger parts of the program are parallelizable.
\Gls{sqs} outperforms dispatch scheduling as it minimizes the movement of shared data and eliminates queuing delay for writers.
At \num{256} threads and \SI{50}{\%} writers, our scheduler improves the throughput by \SI{1.55}{\times}.

Note that \gls{sqs} cannot eliminate queuing delay for every read access, because more than one reader may be awaiting the critical section while a writer holds the mutex.
Once the writer leaves the critical section, it has to resume more than one waiting tasks.
\Gls{sqs} immediately resumes one of the waiters inline and schedules all other waiters on other threads.
Thus, \gls{sqs} eliminates the queuing delay for only one the waiting readers.
Therefore, with a decreasing writer percentage (\ie many readers to few writers) the throughput of \gls{sqs} and dispatch scheduling converges.
For example, at \num{256} threads and \SI{6.25}{\%} writers, \gls{sqs} only improves the throughput by \SI{1.23}{\times}.

\subsection{CES in other Environments}
\label{sec:sqs_in_other_langs}

In this section, we present quantitative measurements to show that (1) \gls{sqs} is a generalizable concept that can be implemented across programming languages and execution systems, and that (2) our mechanism can improve the performance in these settings.
We implement our benchmark program (\cref{alg:mutex_benchmark_task}) in Java Virtual Threads and boost::fibers.
Then, we compare these baseline measurements to implementations of drop-in-replacements that use \gls{sqs}.
The following paragraphs present our findings.

\paragraph{boost::fiber}

\begin{figure}
  \centering
  \includegraphics{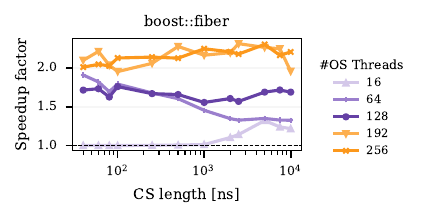}
  \caption{
  Speedup of \gls{sqs} in the boost::fiber::mutex depending on the length of the critical section.
  A thread count larger than 128 spans both NUMA nodes.
  \Description{The speedup of our approach increases with the number of threads used. It is largely independent of the length of the critical section. Speedup is never below 1, i.e., our approach is always as least as performant as the baseline.}
  \label{fig:boost_fiber}
}
\end{figure}

Boost::fibers are an implementation of user\-space threads~\cite{boost_fibers}.
For baseline measurements, we use \lstinline!boost::fibers::mutex!.
We have built a drop-in replacement for this default mutex.
\Cref{fig:boost_fiber} shows the results of our measurements.
Our \gls{sqs} version improves performance by at least \SI{2}{\times} on \num{256} threads.
In particular, our approach is never slower than the baseline.
In absolute numbers, the maximum measured speedup in this measurement is lower compared to our previous measurements on standard C++ coroutines (\cref{fig:mutex_speedup}).
This is because \lstinline!boost::fibers! are stackful coroutines.
Thus, switching the context has a higher overhead, limiting the maximum speedup.
Still, reducing the queuing delay improves throughput despite the overhead.
This highlights that cooperatively scheduled systems regardless of their stack management should not shy away from frequent task switches.
The benefits are unlikely to be outweighed by the overhead of a context switch in the userspace.

\paragraph{Java Virtual Threads}

\begin{figure}
  \centering
  \includegraphics{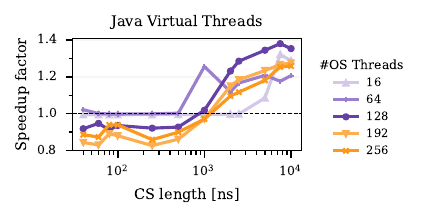}
  \caption{
  Speedup of \gls{sqs} compared to Java's ReentrantLock on Java Virtual Threads depending on the length of the critical section.
  A thread count larger than 128 spans both NUMA nodes.
  }
  \Description{Speedup of our approach over the baseline depends on the number of threads and the length of the critical section. With many threads and short critical sections, our approach degrades performance. For thread counts smaller than 128, our approach makes no difference compared to the baseline on short critical sections. Our approach is consistently outperforming the baseline in long critical sections.}
  \label{fig:java_vthreads}
\end{figure}

Java Virtual Threads are a light-weight thread abstraction implemented with continuations~\cite{Pufek2020, Beronic2021, java_virtual_threads, Beronic2022, Rosa2023}.
They are transparent compared to Java's OS (platform) threads and can be used in legacy applications with near-zero modifications required.
For baseline measurements, we use Java's \lstinline!Reentrant!\-\lstinline!Lock! that schedules Virtual Threads.
We compare this baseline to our proof-of-concept adaption of Virtual Threads and the implementation of a \gls{sqs} mutex.
\Cref{fig:java_vthreads} shows the results of our measurements.
For small critical sections, the overhead required to delegate the critical section with Java's continuations is larger than the benefits of delegation.
Hence, critical sections shorter than \SI{1}{\micro\second} suffer a performance penalty of \SIrange{6}{18}{\%} when using more than \num{64} \gls{os}-level threads.
However, for longer critical sections (\ie >\SI{1}{\micro\second}), the benefits of delegation start to outweigh its overhead.
\Gls{sqs} improves the overall throughput by up to \SI{37}{\%} (\num{128} threads at \SI{7.5}{\micro\second}).
We strongly believe that a well-optimized implementation of \gls{sqs} can reduce the incurred penalty at small critical sections and reap the benefits of improved performance for longer critical sections.
This requires modifications in the JVM, thus, we leave this optimization as future work.

\subsection{Application Benchmarks}
\label{sec:application_benchmarks}

\begin{figure}
  \centering
  \includegraphics{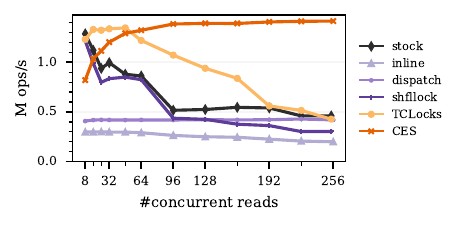}
  \caption{LevelDB readrandom benchmark.}
  \Description{Stock, shfllock and TClocks experience slowdowns with increasing thread counts as they do not optimize parallelism in the system. Our approach increases throughput, outperforming all other implementations.}
  \label{fig:lvldb}
\end{figure}

To analyze the performance on the application scale, we use the popular LevelDB readrandom benchmark and compare stock to shfllock~\cite{kashyap2019scalable}, TCLocks~\cite{tclocks} and our \gls{sqs} mutex.
In the \gls{sqs} version, we have modified only the mutex implementation of LevelDB to use \gls{sqs}.
All other functionality remains the same.
In particular, LevelDB still uses blocking file IO.
Hence, any performance differences are solely due to the use of different synchronization primitives.

\Cref{fig:lvldb} shows the results of our benchmark on the two-socket machine%
\footnote{
    The virtualization layer required to run TCLocks has an implementation limitation at a maximum of 255 mapped processing units.
    Oversubscription with TCLocks leads to system crashes.
    Thus, the measurements shown here are only up to 254 threads.
}.
With up to \num{48} concurrent read operations, \gls{sqs} exhibits a reduced throughput because contention is relatively low.
However, above \num{64} concurrent read operations, \gls{sqs} outperforms all state-of-the-art mutex designs.
Dispatch scheduling incurs queuing delay, which prolongs the critical path (we discuss this further in \cref{sec:impact_queuing_delay}).
TCLocks avoid queuing delay through delegation, but do not maximize parallelism in the system.
This is because waiters in TCLocks use a combination of busy-waiting and kernel-level context switches, limiting parallel throughput.
In contrast, \gls{sqs} eliminates queuing delay and never incurs busy waiting which ensures that parallelizable workloads are executed on all available threads.
Therefore, at 256 concurrent operations, we achieve \SI{3.3}{\times} the throughput of TCLocks (\SI{0.427}{\mega\ops\per\second} and \SI{1.41}{\mega\ops\per\second} respectively).

\subsection{Impact of Queuing Delay}
\label{sec:impact_queuing_delay}

\begin{figure}
  \centering
  \includegraphics{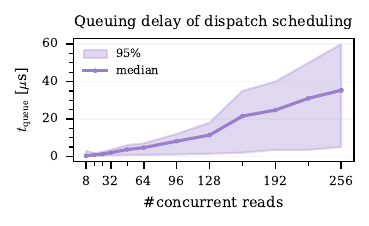}
  \caption{Queuing delay of dispatch scheduling in the LevelDB readrandom benchmark.}
  \Description{}
  \label{fig:lvldb_queuing}
\end{figure}

Finally, we measure the queuing delay of dispatch scheduling to highlight that it is a limiting factor for the application's throughput.
We modify the implementation of the dispatch scheduling mutex in LevelDB and measure the time between a thread \lstinline!schedule()!-ing the next waiter $T$ and this task $T$ entering the critical section (\ie $T$ returns from its call to \lstinline!lock!).
\Cref{fig:lvldb_queuing} shows the results of our measurements.
As expected, the more concurrent reads the system executes concurrently, the longer the ready queues become.
Hence, the queuing delay increases with the number of concurrent reads.
It is striking, how long a task executing the critical section actually resides in the ready queue.
At \num{256} concurrent reads, the median measured queuing delay is \SI{35.5}{\micro\second}, which is two orders of magnitude larger than the median execution time inside of the critical section (\SI{319}{\ns}).

\section{Conclusion}
\label{sec:conclusion}

Cooperative scheduling is experiencing a rennaissance as modern applications once again use coroutines and other userspace threads.
This paper analyzed the implementation of synchronization primitives of cooperatively scheduled environments in modern languages such as C++, Java, or Kotlin.
We found that state-of-the-art implementations fall into one of two categories, where each has substantial shortcomings:
Inline scheduling can result in the collapse of parallelism, where trivially parallelizable tasks end up executing on a single thread.
Dispatch scheduling can result in queuing delay, which describes the time when no task is in the critical section even though at least one task is waiting to enter.
Queuing delay arises if the next lock awaiter is not resumed immediately when unlocking the mutex, but instead is moved to the ready queue of the executor to be resumed at a later point in time.
Hereby, this paper reevaluates the wide-spread assumption that it is okay for locking tasks to wait in the ready queue as long as the process is still getting other work done.
While it feels intuitively right that high utilization corresponds to high throughput, our measurements show that this is not necessarily true.
Specifically, we find that the order of execution is important.
Our main point is that tasks executing sequential parts of the program (i.e., critical sections) should not be delayed and their execution should have priority over parallelizable work.

We present a novel scheduling approach for synchronization in the userspace called \Gls{sqs}, that avoids queuing the locking task's continuation.
The trade-off is that, with \gls{sqs}, the unlocking task's continuation might wait in the ready queue if there is a high workload.
Our evaluation shows that this trade-off is worth it because we shorten the application's critical path by giving priority to the task that executes a critical section instead of the task that has left the critical section.
Our measurements show that \gls{sqs} consistently achieves considerably higher throughput compared to state-of-the-art cooperative scheduling approaches.
In mutex microbenchmarks, we achieve \SI{8.1}{\times} higher throughput compared to dispatch scheduling.
Moreover, on an application-scale, \gls{sqs} improves the performance of the LevelDB benchmark by more than \SI{3}{\times} compared to TCLocks, a state-of-the-art delegation-style lock.
\Gls{sqs} is language-agnostic and can be implemented in any language or asynchronous library.
To show the general applicability of our approach, we have integrated \gls{sqs} in C++20 Coroutines, Java Virtual Threads and boost::fibers, achieving promising performance uplifts.
Hence, the wide-spread adoption of \gls{sqs} could prove beneficial for a wide range of languages, libraries and parallel applications alike.

\bibliographystyle{ACM-Reference-Format}
\bibliography{bibliography.bib}

\end{document}